\documentclass[final,3p,times]{elsarticle}

\usepackage{graphicx}
\usepackage{float}
\usepackage[dvipsnames]{xcolor}
\usepackage{amssymb}
\usepackage{physics}
\usepackage{bm}
\usepackage{cellspace,tabularx}
\usepackage{mathtools}
\usepackage[colorlinks=true,citecolor=blue,urlcolor=blue]{hyperref}
\DeclareUnicodeCharacter{2009}{\,} 
\biboptions{sort&compress}

\date{\today}
\begin{document}

\begin{frontmatter}

\title{Extracting unconventional spin texture in two dimensional topological crystalline insulator bismuthene via tuning bulk-edge interactions
}

\journal{XXX}

%
\affiliation[1]{organization={Department of Materials Science and Engineering, Monash University}, country={Australia}}
\affiliation[2]{organization={School of Physics and Astronomy, Monash University}, country={Australia}}
\affiliation[3]{organization={ARC Centre of Excellence in Future Low-Energy Electronics Technologies (FLEET)}, country={Australia}}

\author[1,3]{Yuefeng Yin\corref{cor1}\fnref{cor2}}
\ead{yuefeng.yin@monash.edu}
\cortext[cor1]{Corresponding authors.}
\fntext[cor2]{These authors contribute equally to this work (Y.Y. and C.W).}
\author[1,3]{Chutian Wang\fnref{cor2}}
\author[2,3]{Michael S. Fuhrer}
\author[1,3]{Nikhil~V.~Medhekar\corref{cor1}}
\ead{nikhil.medhekar@monash.edu}




\begin{abstract}
Tuning the interaction between the bulk and edge states of topological materials is a powerful tool for manipulating edge transport behavior, opening up exciting opportunities for novel electronic and spintronic applications. 
This approach is particularly suited to topological crystalline insulators (TCI), a class of topologically nontrivial compounds that are endowed with multiple degrees of topological protection.
In this study, we investigate how bulk-edge interactions can influence the edge transport in planar bismuthene, a TCI with metallic edge states protected by in-plane mirror symmetry, using first principles calculations and symmetrized Wannier tight-binding models. 
By exploring the impact of various perturbation effects, such as device size, substrate potentials, and applied transverse electric field, we examine the evolution of the electronic structure and edge transport in planar bismuthene.
Our findings demonstrate that the TCI states of planar bismuthene can be engineered to exhibit either a gapped or conducting unconventional helical spin texture via a combination of substrate and electric field effects. Furthermore, under strong electric fields, the edge states can be stabilized through a delicate control of the bulk-edge interactions. These results open up new directions for discovering novel spin transport patterns in topological materials and provide critical insights for the fabrication of topological spintronic devices.
\end{abstract}




\end{frontmatter}
\section{Introduction}

Bulk-edge correspondence, which rigorously links the electronic transport at the boundaries to the bulk electronic structure, is a fundamental concept in topological physics \cite{RevModPhys.82.3045,RevModPhys.88.021004}. 
A one-to-one relation between a bulk-defined topological invariant (e.g. Chern number) and the electronic structure of the edge (e.g. number of edge bands) established by the bulk-edge correspondence has been widely applied to study topologically nontrivial phases such as topological insulators (TIs) \cite{zhang2010crossover}, topological semimetals (TSMs) \cite{PhysRevB.103.195310} and topological crystalline insulators (TCIs) \cite{PhysRevLett.106.106802}.
In practice, however, the edge band dispersion cannot always be inferred from the bulk band topology since the edge electronic structure can be sensitive to local conditions such as doping and perturbations due to a substrate \cite{kotta2020spectromicroscopic,PhysRevB.102.115402,PhysRevB.104.L041102,Rienks2019}. 
This observation does not contradict the bulk-edge correspondence because the existence of the edge bands is protected by the symmetry in the bulk; however, these edge bands can be manipulated by the changes in the local environment. \cite{kotta2020spectromicroscopic,PhysRevB.95.205403,nakamura2022impact,nakamura2020direct,PhysRevB.86.205127}.
This capability of modifying the edge states can be critical for utilizing topologically nontrivial materials in electronic and spintronic devices, which essentially require versatile control over edge transport modes 
\cite{liu2020tunable,colomes2018antichiral,doi:10.1021/acs.nanolett.1c00378}.

Previous studies have revealed that the interplay between the bulk and edge states can either be enhanced or attenuated depending on local physical conditions, and can lead to bulk/edge realignment and the associated emergence of novel physical phenomena in topological materials \cite{doi:10.1021/acs.nanolett.6b04299}. 
In particular, for spin polarized edge states, the bulk-edge correspondence only guarantees the existence of a limited set of spin polarization modes at the boundary (e.g. chiral and helical) \cite{PhysRevLett.110.046404,drozdov2014one,PhysRevLett.115.217601}, but unique spin textures defying the conventional spin distribution can be achieved by tuning the band alignment between the bulk and edge bands \cite{PhysRevB.97.134402,doi:10.1021/acs.nanolett.6b04299,yin2019selective}.
Despite these promising leads, achieving flexible and delicate control over the edge states via bulk-edge interaction without disrupting the bulk band topology remains a challenge. Current approaches typically rely on irreversible modification of the edge state (for example, via local doping) that results in non-switchable changes in the edge states \cite{PhysRevB.98.165146}. 
In order to utilize topologically nontrivial materials in electronic and spintronic applications, the device would need to allow for a convenient way to transmit different types of electronic and spin signals while retaining the ability to switch between topological phases \cite{collins2018electric,adma-202005698}. 
In contrast to relying on a crossover between edge and bulk inhomogeneities to manipulate bulk-edge interactions, utilizing external means such as strain and electric/magnetic field to influence the local environment offers a favorable approach to deliver a non-invasive, flexible and switchable control knob to tune edge transport while retaining the topological protection \cite{huang2019proximity,PhysRevB.95.205403,PhysRevB.97.134402}.
For example, it has been shown that van der Waals heterostructures formed by topological materials supported on substrates can display exotic edge spin texture as a consequence of the induced charge transfer by an external electric field \cite{PhysRevB.97.134402}. 
As increasing attention is being drawn on developing efficient strategies to tune edge transport in topological materials, it is critical to understand the mechanisms underlying external control of the interactions between bulk and edge bands \cite{adma-202005698}.

Among the various types of topological phases, TCIs offer a unique pathway for tuning of edge states since the topological protection in a TCI arises due to specific crystalline symmetries \cite{PhysRevLett.106.106802,PhysRevB.88.241303}. 
In particular, for 2D thin films, the TCI phase is defined by a mirror symmetry-related topological index, known as the mirror Chern number. 
Planar bismuthene is one of such 2D TCIs classified by the mirror Chern number $C_M$=2, corresponding to the two branches of bands at the edge \cite{hsu2016two}.
The fragile nature of TCI has made the planar bismuthene a promising target to manipulate edge states by external fields and substrates \cite{nouri2018topological,nouri2020analysis,Li2021NJP,PhysRevB.88.241303,hsieh2012topological}.
Here we have conducted a systematic investigation on approaches for tuning the bulk-edge interactions in TCI planar bismuthene using basis-expanded symmetrized tight-binding Hamiltonians derived from first principles calculations. 
We start by examining the inherent structural factors such as edge configurations and finite size effects to reveal the possibility of enhancing the bulk-edge interactions via altering the hybridization between bulk and edge states.
The edge degeneracies solely protected by mirror symmetry provide flexibility over the control of the alignments between the bulk and edge bands.
We show that with a combination of the substrate potential and the external electric field, we can achieve a delicate extraction of specific edge channels, resulting in either gapped or conducting unconventional helical spin texture. 
These findings highlight the important role of bulk-edge interactions, which provide a new degree of freedom for topological tuning without disrupting the intrinsic topological phase. 
Understanding these interactions can offer key physical insights and guidelines for designing future electronic and spintronic devices.

\section{Methodology}
The electronic structure of bismuth allotropes is calculated using density functional theory (DFT) as implemented in the Vienna \textit{ab initio} Simulation Package (VASP) \cite{KRESSE199615}.
The Perdew-Burke-Ernzehof (PBE) form of the generalized gradient approximation (GGA) is used to describe electron exchange and correlation \cite{PhysRevLett.77.3865}. 
The kinetic energy cutoff for the plane-wave basis set is set to 400 eV. 
We use a 13$\times$ 13 $\times$ 1 and 5$\times$ 1 $\times$ 1 $\Gamma$-centered $k$-point mesh for sampling the Brillouin zone of two-dimensional bismuth allotropes and nanoribbons, respectively.   
All structures are fully relaxed until the ionic forces are smaller than 0.01 eV/\AA. 
A 15 \AA \ vacuum separates the periodic image to avoid any spurious interactions.

\begin{figure*}[htbp]
  \begin{center}
  \includegraphics[scale=0.35]{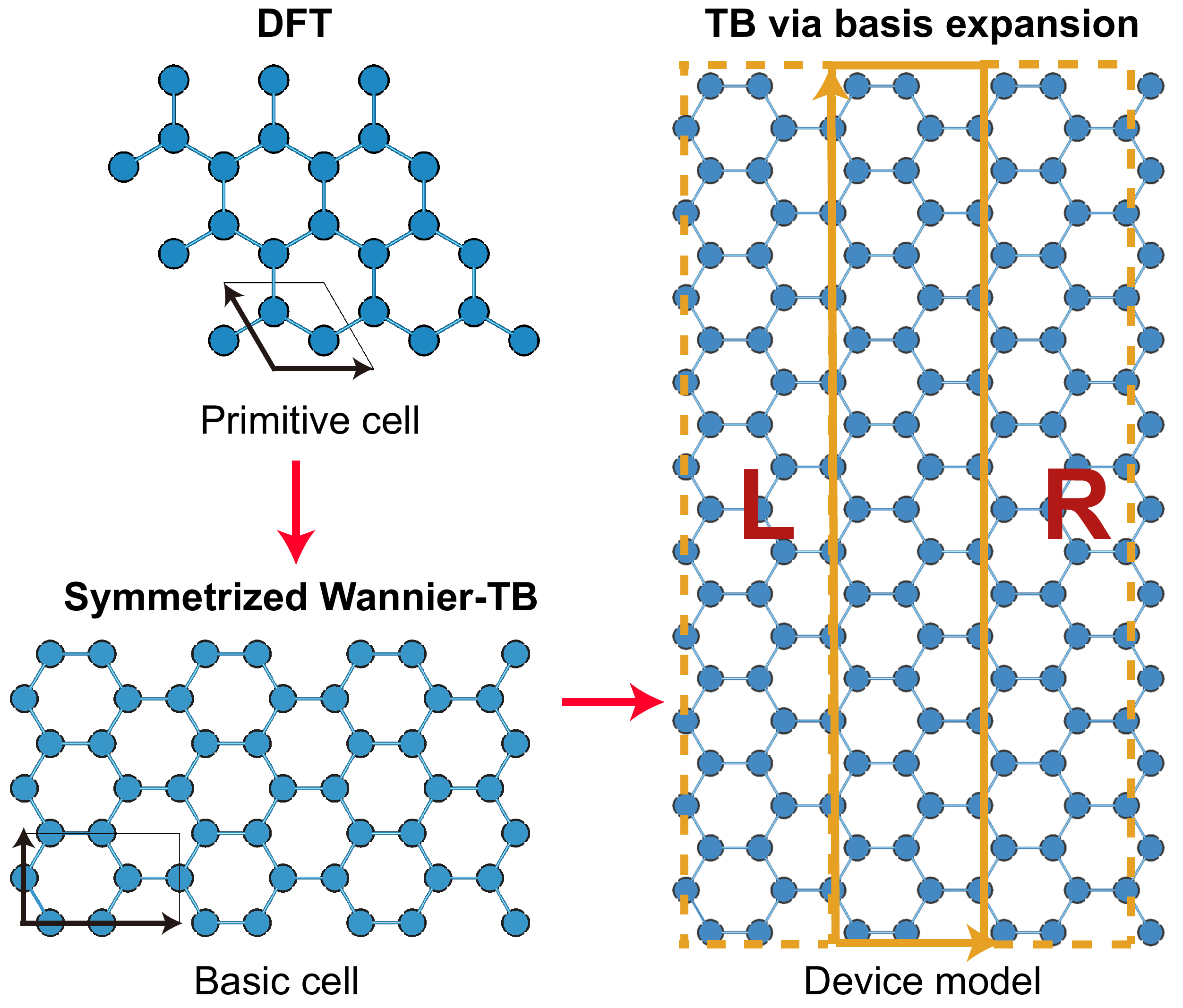}
  \end{center}
  \caption {\small Computational workflow of using expanded TB Hamiltonian for calculating electronic structure and transport in nanoribbons. We start with a DFT calculation on the hexagonal  primitive unit cell, followed by projecting the obtained wavefunctions to localized Wannier basis. The Wannier TB Hamiltonian of the primitive cell is then transformed to that of the orthorhombic unit cell as the construction block of the nanoribbon (referred as the ``basic cell''). The basic cell is finally expanded to form the Hamiltonian of a nanoribbon in device dimensions. }\label{fig1}
  \end{figure*}

Since it is computationally prohibitive to implement first principles calculations on the scale of device dimensions involving a large number of atoms, we have adopted the approach of transforming the plane-wave DFT data into highly localized Wannier functions \cite{mostofi2008wannier90}. 
The Hamiltonian derived from the localized Wannier basis can recover critical electronic and topological features of materials with much less computational cost. 
In our previous work \cite{Li2021NJP}, we have demonstrated that these Wannier-based tight-binding (TB) Hamiltonians can effectively interpret the novel topological phases in two-dimensional allotropes of bismuth. 
Our Wannier-TB models utilize a symmetrization procedure to correct the numerical errors arising during the projection from the DFT Bloch basis. This symmetrization process enhances the agreement between the band dispersions obtained from DFT calculations and those obtained from the Wannier-TB approach.\cite{Li2021NJP} 
Here we generate the TB Hamiltonian of two-dimensional bismuth nanoribbons to calculate the electronic transport in a transistor device setup. 
The TB Hamiltonian of the nanoribbon is restructured and expanded from that of an orthorhombic unit cell (referred as the 'basic cell' in Figure \ref{fig1}). 
The size of the basic cell is determined by the number of nearest-neighbor hoppings considered, ensuring the interaction range does not exceed the nearest neighboring cell along the transmission direction. 
In the case of planar bismuthene, the hoppings up to the second-nearest neighbors are sufficient to reproduce the electronic band structure \cite{Li2021NJP}, leading to a basic cell size of 1 $\times$ 1 as shown in Figure \ref{fig1}. 

The final step of obtaining the Hamiltonian for the device model is to apply a periodic expansion on the Hamiltonian of the basic cell. 
In the device model (right half of Figure \ref{fig1}), the horizontal axis is the transport direction with a periodic boundary, while the vertical axis is the non-periodic direction of confinement. 
The middle cell is the transport channel and adjoining left and right cells (dashed rectangles in Figure \ref{fig1}) along the transport direction act as contacts in a device.
The size of the device is represented by the width of the nanoribbon. 
In a typical calculation for planar bismuthene, a device length of 20 nm is made of 40 basic cells with 480 orbitals, basis expanded from 6 basis in the primitive cell of bismuthene.
Once we have the device model in the form of a sparse TB Hamiltonian expressed in localized Wannier basis, we solve for the transport properties of the system using a non-equilibrium Green's Functions (NEGFs) formalism \cite{KLYMENKO2021107676,liu2021chirality}.

To determine the spin texture of the edge states, we need to obtain the spin operators $\hat{S}_{i}$ ($i=x,y,z$) for the expanded Hamiltonian and then we project the spin operators $\hat{S}_{i}$ onto the edge states subspace $\Psi$ as \cite{PhysRevB.82.045122} :
\begin{equation}
\bra{\Psi}\hat{S}_{x}\ket{\Psi} = S_{x0}\sigma_{x}
\end{equation}
\begin{equation}
\bra{\Psi}\hat{S}_{y}\ket{\Psi} = S_{y0}\sigma_{y}
\end{equation}
\begin{equation}
\bra{\Psi}\hat{S}_{z}\ket{\Psi} = S_{z0}\sigma_{z}
\end{equation}
Here, $\sigma_{x(y,z)}$ are Pauli matrices and the expectation values $S_{x\left(y,z\right)}$ are defined as the spin textures.
In the confined 1D nanoribbon scenario, states with positive/negative $S_z$ polarizations are similar to conventional understanding of ``spin-up''/``spin-down'' components when SOC is absent. In the subsequent sections, we use ``spin-up''/``spin-down'' notions to separate states with positive/negative $S_z$ spin polarizations, providing a better correspondence to experimental results obtained in spin- and angle-resolved photoemission spectroscopy (SARPES) \cite{PhysRevB.94.195134}.

\section{Results and Discussion}

\subsection{Pristine 2D and 1D planar bismuthene}

\begin{figure}[htbp]
\begin{center}
\includegraphics[scale=0.26]{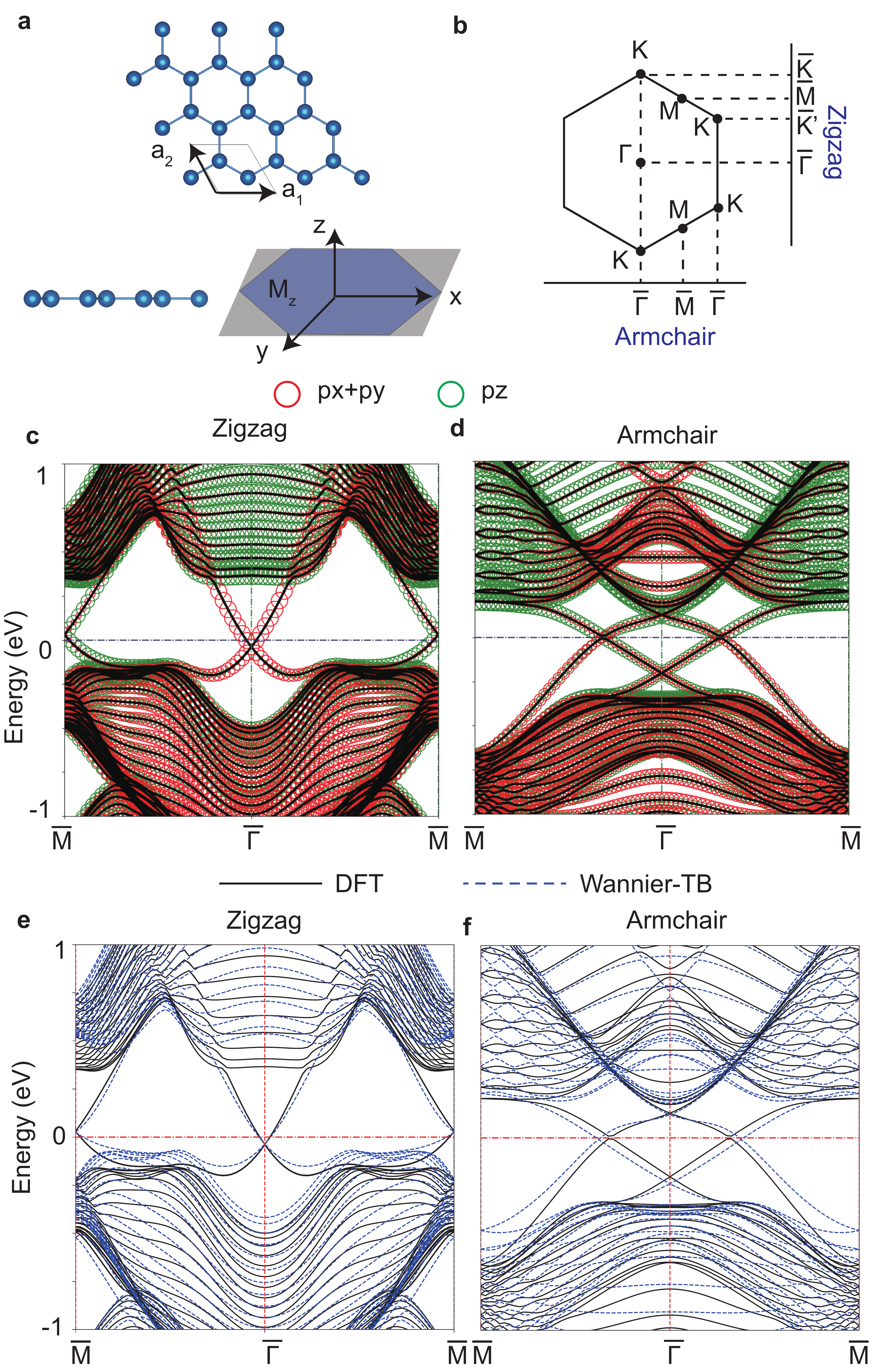}
\end{center}
\caption {\small 
Electronic structure of planar bismuthene nanoribbons. (a) Top and side view of planar bismuthene ($a_1$ and $a_2$ denote the lattice vectors for the primitive cell); $M_z$ is the mirror plane. (b) Projection of 2D Brillouin zone onto the 1D Brillouin zone of zigzag nanoribbon ($k_y$) and armchair nanoribbon ($k_x$). (c-d) DFT band structure of zigzag (5.7 nm wide) and armchair nanoribbon (6.2 nm wide). Red and green colors denote $p_x$+$p_y$ and $p_z$ orbital contributions, respectively. (e-f) Comparison of the DFT band structure (black solid lines) and the bands obtained from TB models (blue dashed lines) for zigzag and armchair nanoribbons. In zigzag nanoribbons, the edge hopping parameters in TB basis are altered to downshift the position of Dirac cones at $\Gamma$ and M to match DFT results.}\label{fig2}
\vspace{\baselineskip}
\end{figure}

We first briefly revisit the electronic properties of 2D and 1D planar bismuthene. The 2D planar bismuthene has a hexagonal crystal structure with an optimized lattice parameter of $a=5.27$ \AA \ as obtained from our DFT calculations (Figure \ref{fig2} (a)). 
The 2D planar bismuthene exhibits a $M_z$ mirror plane (Figure \ref{fig2} (a)), which is crucial for the emergence of the TCI phase \cite{hsu2016two}.
Since the electronic bands near the Fermi level mainly consist of Bi \textit{p} orbitals, we only chose  $p_x$, $p_y$ and $p_z$ orbitals of Bi as projectors to transform the DFT Bloch basis to localized Wannier functions \cite{Li2021NJP}.
After enforcing symmetry constraints and truncating the hopping interaction up to the second-nearest neighbors, we obtain a TB Hamiltonian of 6 atomic orbital basis with 9 independent hopping parameters for the primitive unit cell of planar bismuthene. 
To include spin orbit coupling (SOC), the number of orbitals in the basis is doubled to account for the spin degree of freedom. 
The values of hopping parameters and SOC strength are the same as in our previous work \cite{Li2021NJP}.

Figure \ref{fig2} summarizes key results for the electronic structure of zigzag and armchair bismuthene nanoribbons.
In order to examine the electronic structure of 1D nanoribbons, it is useful to see how the high symmetry points in the 2D Brillouin zone are projected onto the 1D Brillouin zone for different confinement scenarios as shown in Figure \ref{fig2} (b). 
These differences in projections between the zigzag and armchair edge result in contrasting edge dispersion and orbital characters in the band structure of nanoribbons obtained from DFT calculations (Figure \ref{fig2} (c) and (d)).
In the zigzag nanoribbon, the edge states are separated into two distinct branches: a branch with band degeneracy at $\Gamma$ contributed mainly by $p_x$+$p_y$ orbitals, and a branch centering on $M$ contributed by $p_z$ orbitals. The two edge branches intersect near the $K'$ point and vanish into the bulk bands. 
This observation reflects the orbital filtering effect at $\Gamma$ and $M$ in the 2D band structure as reported in our previous investigation \cite{Li2021NJP}.
While for the armchair nanoribbon, because the $K$ and $\Gamma$ states in the 2D Brillouin zone are both projected onto the $\Gamma$ in 1D, the edge orbital character near $\Gamma$ becomes entangled with both $p_x$+$p_y$ and $p_z$ states.
For topologically nontrivial structures, the most important feature is the conducting edge channel in the form of edge band degeneracies within the bulk band gap. In Figure \ref{fig2} (c) and (d), we can see that the location of these edge degeneracies is also different for the zigzag and the armchair edge. The zigzag nanoribbon only features degeneracies at high symmetry point $\Gamma$ and $M$, while the armchair nanoribbon have degeneracies along $\Gamma-M$. These in-between degeneracies are solely protected by the TCI band topology. 
These observations broadly agree with other 2D hexagonal materials that share the same crystal symmetry \cite{araujo2019interplay,hsu2016two}. 
However, the anisotropy in the band character of bismuthene nanoribbons is much stronger and complex than previous examples such as graphene due to the presence of two branches of edge states guaranteed by the bulk-edge correspondence (Mirror Chern number $C_M=2$) \cite{PhysRevB.73.235411,PhysRevB.100.205111}.

As illustrated in Figure \ref{fig2} (e) and (f), our TB models can effectively capture the key features in DFT calculations. 
In our DFT calculations, we have observed small changes in bond lengths in the zigzag edge (bond length reduction of 2.6\% compared to the bulk). 
While such changes are ignored in TB calculations, we can still match the location of the edge degeneracies obtained using TB models to DFT by tuning the on-site potential of $p_x$/$p_y$ and $p_z$ orbitals as demonstrated in Figure \ref{fig2} (e).
Because the structural changes are restricted to the molecular plane, the edge degeneracies are still protected due to the mirror symmetry.
These results also reveal that the subtle changes in edge structure can alter the alignment between the edge and bulk bands without disrupting the topological phase of the system.
In armchair nanoribbons, we have observed negligible structural changes after relaxation using DFT. Consequently, the edge dispersions obtained from DFT calculations and our TB model agree well (Figure \ref{fig2} (f)). 
It should be noted our TB models are computationally more efficient than full-scale DFT methods. This advantage will become more substantial when we investigate the effect of substrate and electric field later.

\subsection{Size effect and edge terminations}

\begin{figure*}[htbp]
\begin{center}
\includegraphics[scale=0.3]{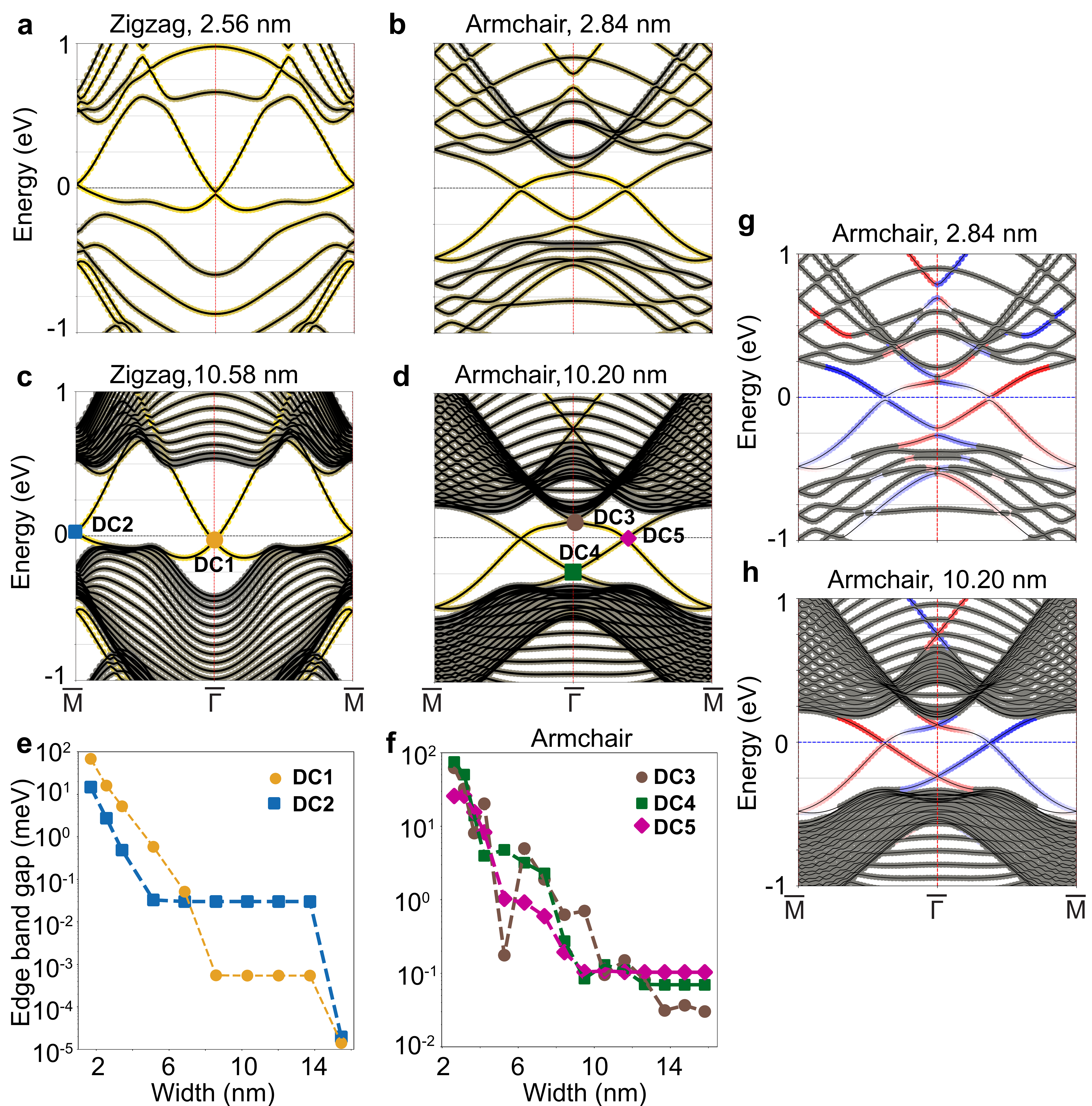}
\end{center}
\caption {\small Evolution of edge band dispersion and spin texture with the width of bismuthene nanoribbons. (a–d) Low energy band structures of a narrow and a wide nanoribbon in zigzag (a, c) and armchair (b, d) configuration.  Yellow and grey circles represent the contribution from the edge states and the bulk states, respectively. 
(e, f) The band gap at various edge band crossing points in a zigzag and an armchair nanoribbon as a function of its width, respectively. The dashed lines mark the threshold for a complete closing of the edge band gap. 
(g,h) Spin-resolved edge band structures in a narrow and a wide armchair nanoribbon, with red (blue) circles indicating spin-up (spin-down) projections in the bands with significant edge contributions. The grey circles on the bands represent contributions from bulk. 
}\label{fig3}
\vspace{\baselineskip}
\end{figure*}

Next we consider how the width of the bismuthene nanoribbon influences its electronic dispersion. 
It is well known that the quantum confinement effect attributed to the inter-edge interactions will induce band gap opening in a narrow nanoribbon \cite{ritter2009influence}.
When the width of a bismuthene nanoribbon increases, the confinement-driven overlap of the edge states is reduced, resulting in the stabilization of the conducting edge. 
We can clearly observe this behavior by comparing the band structures of narrow and wide bismuthene nanoribbons as shown in Figure \ref{fig3} (a-d).
We use the magnitude of the edge band gap as a function of the width of the nanoribbon to quantify the strength of the quantum confinement as shown in Figure \ref{fig3} (e) and (f). 
In the case of zigzag nanoribbons, two edge band crossing points DC1 and DC2 are located at $\Gamma$ and $M$, respectively. 
The edge band gap at both DC1 and DC2 crossing points decays exponentially as the width of the nanoribbon increases (Figure \ref{fig3} (e)).
This trend is consistent with previous studies for other two-dimensional materials \cite{PhysRevB.54.17954,dolui2012electric}.
In the armchair nanoribbon, apart from two edge band crossing points at $\Gamma$ (DC3 and DC4), there is an additional crossing point DC5 along $\Gamma - M$ (Figure \ref{fig3} (f)). 
It can be observed that gaps at DC3 and DC4 crossing points in the armchair nanoribbon oscillate moderately with the width of the nanoribbon, which can be attributed to the proximity of these band crossing points to the bulk electronic states. 
This trend is consistent with a previous report on the role of finite size effects in TCI \cite{PhysRevB.90.045309}.
For both armchair and zigzag nanoribbons, it is observed that the edge band gap maintains a relatively low and constant magnitude when the width of the nanoribbon exceeds 8 nm ($<$0.1 meV), indicating a minimal influence of the finite size effect.

TCIs are known to have spin-filtered edge states, a feature similar to the helical edge states in topological insulators \cite{liu2014spin}. 
For planar bismuthene, two branches of such spin-filtered bands exist on each edge (Figure \ref{fig3} (g) and (h) for the armchair nanoribbons). This observation is consistent with the previously reported results on TCIs (e.g. SnTe) sharing the same mirror Chern number $|C_{M}|=2$ \cite{liu2014spin}. 
By comparing the spin-resolved band dispersions in the narrow and wide armchair nanoribbons as shown in Figure \ref{fig3} (g) and (h), it can be observed that the spin polarization remains unchanged even when band degeneracies are lifted due to the quantum confinement.  This also echoes previous reports of the robustness of spin transport along the edge dispersion of TCI against perturbation effects such as temperature and external fields \cite{liu2014spin,PhysRevB.98.165146}. This spin-filtered behavior can be further modified by application of external fields to create an effective spin switch, as external fields can lead to in-gap tunneling of spin currents between edges \cite{doi:10.1126/science.aah6233}. 
Next, we explore how the edge states of planar bismuthene can be tuned via substrates and external factors such as strain and electric fields, and focus on the interactions between bulk and edge bands that can lead to new electronic transport patterns. 
 
\subsection{Effect of the substrate potential}

\begin{figure}[htbp]
\begin{center}
\includegraphics[scale=0.18]{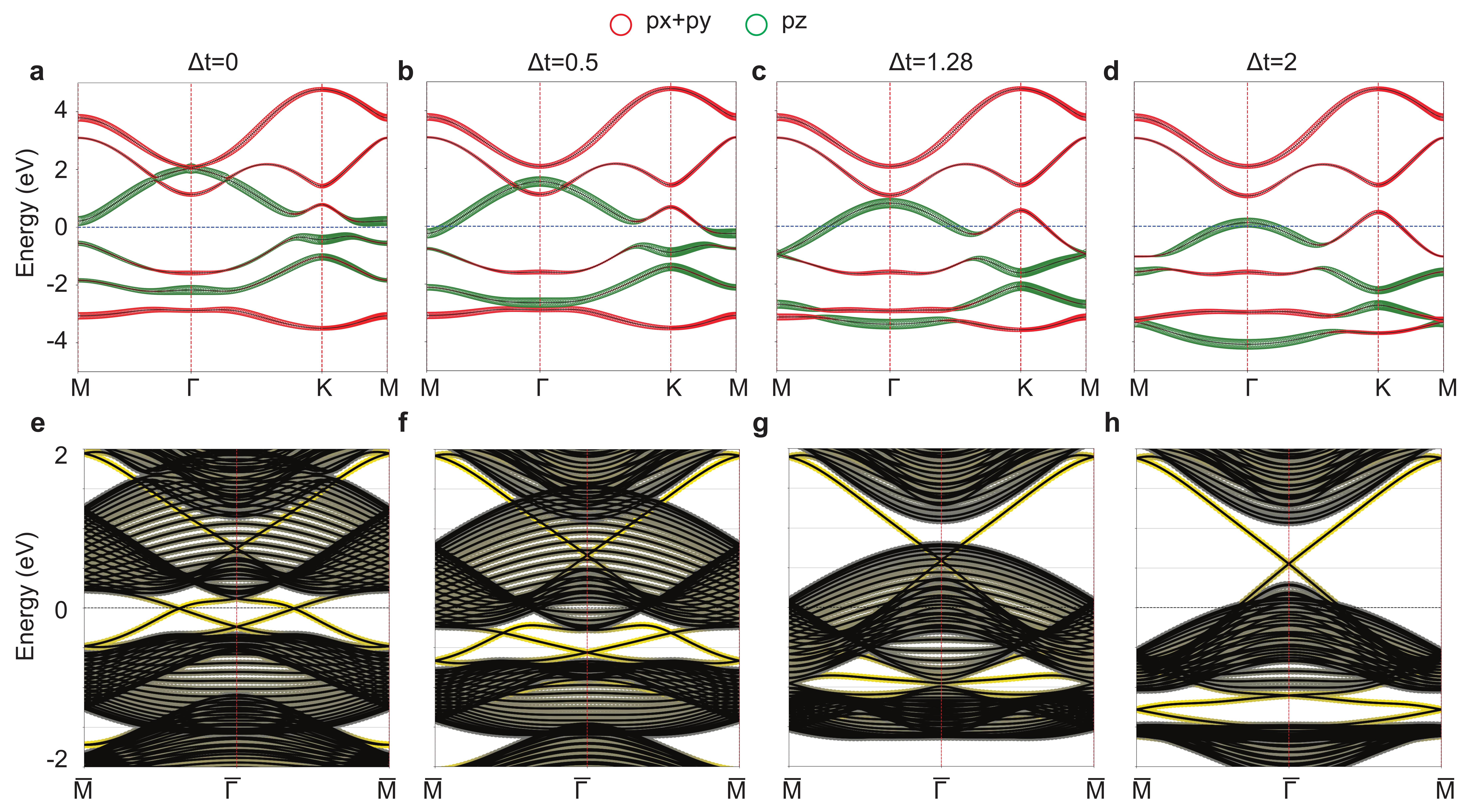}
\end{center}
\caption {\small  Evolution of the electronic band structure with a small, uniform on-site potential $\Delta t$ (in eV) on $p_z$ orbitals for 2D planar bismsuthene (a-d) and bismuthene armchair nanoribbon (e-h). Red and green circles in (a) indicate contributions of $p_{x}+p_{y}$, and $p_{z}$ orbitals, while grey and yellow circles in (b) denote contribution from the bulk and the armchair edge, respectively.  }\label{fig4}
\vspace{\baselineskip}
\end{figure}

In any practical design of the electronic device using 2D topological materials, the supporting substrate provides a useful knob to exert control over the edge dispersions.  For a TCI, the choice of the substrate is especially important as the crystalline symmetry that protects the conducting the edge states 
can be sensitive to the interactions between the substrate and the thin film \cite{PhysRevB.88.241303}. 
Strong substrates such as SiC can form covalent bonds with the planar bismuthene, inducing a topological phase transition from a TCI to a TI \cite{reis2017bismuthene}. 
To preserve a TCI phase, an ideal supporting substrate should have a weak interaction (e.g. van der Waals interaction) with the supported 2D topological layer.
To simulate the influence of a weakly-interacting substrate on the band dispersion in bismuthene, we use a uniform onsite potential $\Delta t$ on $p_z$ orbitals in our Wannier-based tight-binding model, which quantifies the strength of binding with the substrate and increases the anisotropy and hybridization between $p_z$ orbital and $p_x$/$p_y$ orbitals.
This approach has been explored in previous studies to assess the interaction between van der Waals-bonded thin films and substrates  \cite{PhysRevB.87.245408,PhysRevB.105.245145,Attig_2021}. For instance, in the case of graphene supported on hexagonal boron nitride (h-BN), the substrate potential exerted by h-BN on graphene’s $p_z$ orbitals has been estimated to be 0.8 eV \cite{PhysRevB.87.245408}. Weak perturbations resulting from van der Waals effect in 2H-NbS2 can lead to changes in the substrate potential in the range of vary from 0.3 to 1.4 eV o, contingent upon the separation distance between the film and the substrate, as well as the extent of substrate misfit.\cite{PhysRevB.92.205108}
Here we focus on the change in the electronic structure of the armchair edge since it demonstrates a much more hybridized and entangled edge dispersion compared to the zigzag case. In addition, the degeneracies in the edge band structure of the armchair nanoribbon are supported only by the mirror symmetry, allowing the armchair nanoribbons to be a good candidate for exploring the influence of substrates. 

Figure \ref{fig4} shows the band structures of 2D bismuthene and 1D armchair bismuthene nanoribbons with different onsite potentials on $p_z$ orbitals.
In the case of free-standing 2D bismuthene ($\Delta t=0$), the $p_z$ orbitals are concentrated in the conduction band along $\Gamma - M$ and $\Gamma - K$ (Figure \ref{fig4} (a)). 
The additional on-site potential shifts these $p_z$ states down and causes the band gap to close. 
The gap between the original valence band and the conduction band at $\Delta t=0$ eV is fully closed at $\Delta t=1.28$ eV at $M$ (Figure \ref{fig4} (c)). 
When the $p_z$ on-site potential approaches 2.0 eV, the gap reopens, but the order of band character at $M$ is exchanged. 
As the original valence band has now been pushed deep below the Fermi level, the low-energy region is now left with the 4th and 5th bands, which are composed of $p_x+p_y$ states with a gap of 0.94 eV at K.

We can further observe the effect of the on-site potential to $p_z$ orbitals in the band structure of 1D armchair nanoribbons (Figure \ref{fig4} (e-h)). 
For small potentials ($\Delta t < 0.5$ eV), the edge states display the character of a TCI, with  edge band degeneracies being pushed closer to the bulk band region. 
After the original bulk band gap is closed, the edge bands are fully buried in the bulk bands (Figure \ref{fig4} (g)). Upon further increasing $\Delta t$, the buried edge bands begin to emerge and the top branch of an edge band is clearly seen along $\Gamma–M$. 
Finally, when the added potential reaches 2.0 eV, the buried edge bands become completely disentangled from the bulk bands, reflecting the band order change as seen in Figure \ref{fig4} (d).
This observation is in agreement with the feature observed in the TI phase in bismuthene covalently bonded with the SiC(0001) substrates due to a substrate-orbital-filtering effect \cite{reis2017bismuthene,doi:10.1021/acsnano.1c09592}.
These results give us a clear hint on how substrates can allow for tuning of the edge dispersions in 2D TCIs by modifying the interplay between bulk and edge states.

\subsection{Interplay between edge and bulk states via external electric field}

\begin{figure}[htbp]
  \begin{center}
  \includegraphics[scale=0.25]{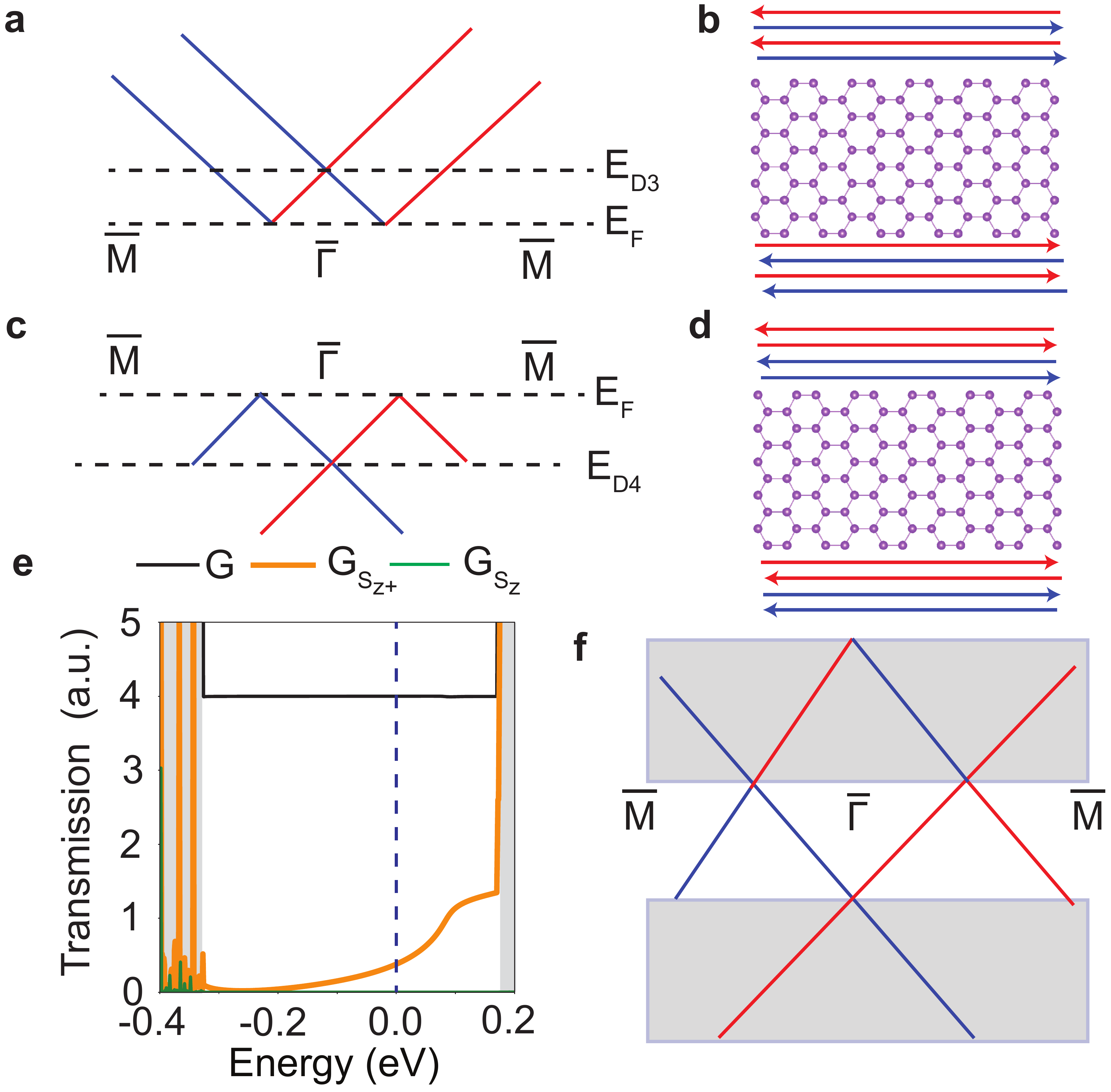}
  \end{center}
  \caption {\small Schematics (a-d) summarize the differences in spin textures in bismuthene armchair nanoribbon. The red and blue lines denote spin-up and spin-down edge states, respectively. 
  (a-b) show the spin texture and the edge transport above the Fermi level, while (c-d) show the spin texture and the edge transport below the Fermi level. The Fermi level is also where the edge degeneracies protected by the mirror symmetry are located.
  (e) shows the total conductance ($G$), spin-up transmission probability ($G_{S_{z+}}$) and spin conductance ($G_{S_{z}}$) for the edge dispersion in the pristine armchair nanoribbon as depicted in Figure \ref{fig3} (h).
  (f) shows how the spin texture below the Fermi level can be extracted via burying other edge states into the bulk. 
  Grey shaded area represents the bulk region.
  }\label{fig6}
  \vspace{\baselineskip}
  \end{figure}

In addition to substrates, external electric field provides a useful approach to manipulate the topological and electronic properties of the 2D thin film \cite{bihlmayer2022rashba,wang2018electric}. For 2D TCI, the external electric field is generally regarded as a switch to break the symmetry-protected edge states \cite{PhysRevB.98.165146} and can lead to spin splitting and band shifts. %
Here we first investigate the impact of electric field on the band dispersion in free-standing planar bismuthene.
Our DFT calculations (see Supplementary Figure S1) show that the band structure of the free-standing planar bismuthene layer  exhibits no significant energy level shifts under the electric field except for a  small band splitting. This is in contrast to multilayer structures or buckled lattices where the applied electric field can result in considerable band shifts by inducing a potential difference between different layers or sublattices \cite{PhysRevB.91.235145,PhysRevB.98.035106}. 
The observed band splitting in planar bismuthene can be modeled via a Rashba term in our TB model. We can estimate the relationship between the Rashba coefficient and the electric field strength by fitting the TB band dispersion to the DFT band structure as $E_{eff}= 11.5\lambda_{R}$.
This Rashba effect induces a subtle mixing of the bulk and edge electronic dispersion and a small band splitting in planar bismuthene, which is not strong enough to lead to changes in band ordering, indicating that the intrinsic nature of the band topology remains unchanged. 
These observations motivate us to investigate a combined effect both the on-site energy shift to induce orbital anisotropy (via a weak substrate effect) and Rashba interactions (via electric field) to tune the edge transport in 1D TCI nanoribbons. 
For this purpose, we focus on the bismuthene armchair nanoribbon since the armchair edge contains degeneracies only protected by the mirror symmetry, which is susceptible to the influence of symmetry-breaking electric field.

Figure \ref{fig3} (h) shows that the spin texture of bismuthene armchair nanoribbons is helical, similar to that observed in other topological materials with the quantum spin Hall effect. However, we have observed that the helicity is not uniform across the gapped region as shown in Figure \ref{fig6}. Above the Fermi level (corresponding to the energy level of the mirror-symmetry protected degeneracy $E_{DC5}$), the edge states exhibit a W-shaped conventional helical spin texture (Figure \ref{fig6} (a)). This implies that the left-moving and right-moving electrons on the same edge possess different spin polarizations, as shown in Figure \ref{fig6} (b).
In contrast, when we consider the states below the Fermi level, we find  an unconventional spin texture  between the Fermi level and the energy level of DC4. The spin polarization near $\overline{M}$ flips signs (Figure \ref{fig6} (c)). This implies that on each edge, the spin-up or spin-down electrons are no longer confined to a fixed direction (left or right-moving) as dictated by the quantum spin Hall effect (Figure \ref{fig6} (d)). This behavior was previously suggested only when the inter-edge interactions were strong and the edge states could scatter via tunneling\cite{stuhler2022effective,PhysRevLett.102.096806}. However, spin-flip scattering still requires a large momentum transfer, and therefore may be strongly suppressed for long-wavelength disorder. This behavior might be advantageous for pure spin transport. In this spin texture, the nanoribbon can support both pure charge current without net spin transport, and pure spin currents without net charge transport; both impossible in the helical spin texture of Figure \ref{fig6} (a). 

In an ideal two-terminal setup, both the helical spin texture (Figure \ref{fig6} (a)) and the flipped transport pattern (Figure \ref{fig6} (c)) exhibit a spin-neutral behavior, resulting in a net spin current of zero in the absence of magnetic disorders \cite{PhysRevLett.95.226801}.
However, compared with the W-shaped transport pattern,  the conductance for individual spin components (spin-up/spin-down) in the M-shaped transport pattern shown in Figure \ref{fig6} (c) would be much reduced due to the presence of two branches of canceling currents. This difference in spin-conductance between W-shaped current and M-shaped current can be observed in the edge conductance curves of pristine armchair nanoribbons in Figure \ref{fig6} (e). The edge conductance is quantized at 4 $e^2/\hbar$, matching the presence of two branches of conducting currents, while the net spin conductance is zero as the transmission probability for spin-up and spin-down currents are identical. The spin conductance for the spin-up (spin-down) component is non-zero and shows a fluctuating trend in the edge region.

We observe that the W-shaped edge bands near the bulk conduction band exhibit significantly larger spin conductance compared to the M-shaped region around and below the Fermi level, due to the canceling effect of M-shaped edge currents. Under the influence of perturbation effects such as field and substrate potentials, the conductance for individual spin components can be further adjusted by tuning the amount of spin polarization in each pair of spin currents. By using these tuning strategies in combination, we can create a switch between spin currents of different magnitudes and orientations in the conducting edges, especially for the M-shaped transport pattern, which allows canceling and reversing the spin currents.
In the following, we will focus on how to use the bulk-edge interactions to isolate this unconventional spin texture by burying other helical states into the bulk. For example, we can achieve isolation of the transport pattern shown in Figure \ref{fig6} (c) by modifying the edge dispersions to the one shown in Figure \ref{fig6} (f).

\begin{figure}[htbp]
  \begin{center}
  \includegraphics[scale=0.18]{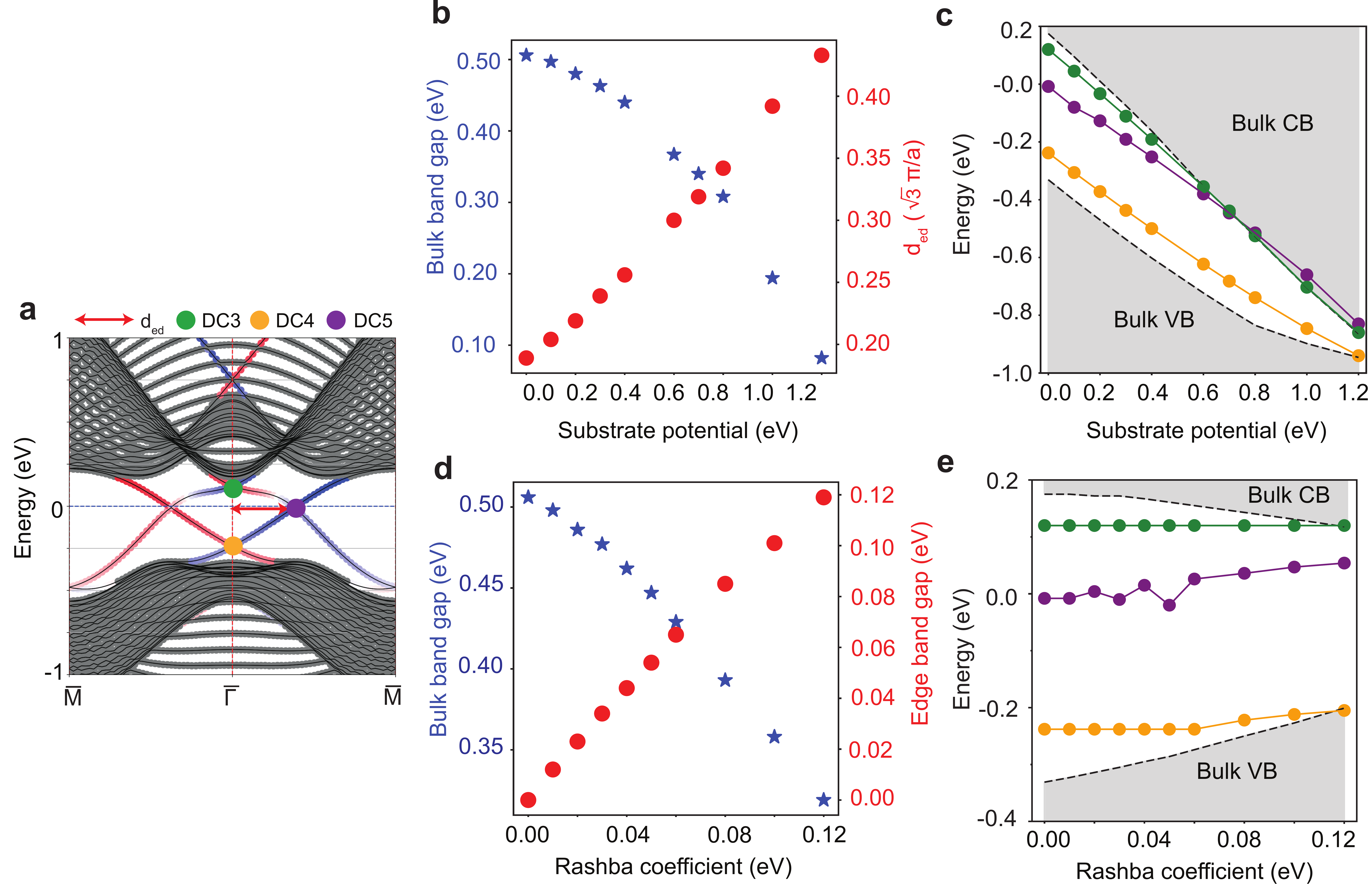}
  \end{center}
  \caption {\small Effect of the substrate potential and the electric field on the bulk-edge band alignments in a 9 nm wide armchair bismuthene nanoribbon. (a) A schematic for describing the shifts in various degeneracy points with the substrate potentials and electric field.  
  (b-c) Effect of the substrate potential on the bulk band gap and the distance of the mirror symmetry point DC5 to $\Gamma$ (as indicated by red arrows in (a)).
  (d-e) Effect of the external electric field on the bulk band gap and the edge band gap at the degeneracy point DC5. 
  (c,e) Shifts in various degeneracy points, degeneracies disappear into the bulk region with increasing substrate potential (c) and external electric field (e). 
  Note that if DCs are gapped by a strong substrate potential or electric field, the energy at the mid-gap position is assumed to represent the degeneracies.
  }\label{fig7}
  \vspace{\baselineskip}
  \end{figure}

Figure \ref{fig7} summarizes the effect of substrate potential and electric field on the bulk-edge band alignments.
With increasing substrate potential, the bulk band gap is reduced and the edge band degeneracy at DC5 is pushed away from $\Gamma$ and towards $M$, as illustrated by the red arrow in \ref{fig7} (a) with the data presented in \ref{fig7} (b). The bulk band gap reduces rapidly once the substrate potential reaches 0.8 eV, while the distance of the the edge degeneracies DC5 to $\Gamma$ shows a nearly linear response.
Figure \ref{fig7} (c) shows how all edge degeneracies are gradually buried into the bulk conduction or valance bands with increasing substrate potential. It is worth noting that the degeneracy at DC3 disappears into bulk region at a low substrate potential of 0.6 eV. 
When all degeneracies are buried, the edge bands resemble a ``M-shaped'' configuration, qualitatively similar to the schematic Figure \ref{fig6} (b).
In contrast to the effect of the substrate potential, the most notable effect of the external electric field is the breaking of the mirror edge degeneracies at DC5 and introduction of a small edge band gap (red circles in Figure \ref{fig7} (d)).
With increasing electric field the bulk band gap also reduces while the edge band gap increases as shown in Figure \ref{fig7} (d)). Overall, the reduction in the bulk band gap with external electric field is smaller compared to that due to the substrate potential (Figure \ref{fig7} (b)).
Similar to what is observed with increasing substrate potential, all edge degeneracies also move close to the bulk region as the electric field increases (Figure \ref{fig7} (e)).

\begin{figure}[htbp]
  \begin{center}
  \includegraphics[scale=0.19]{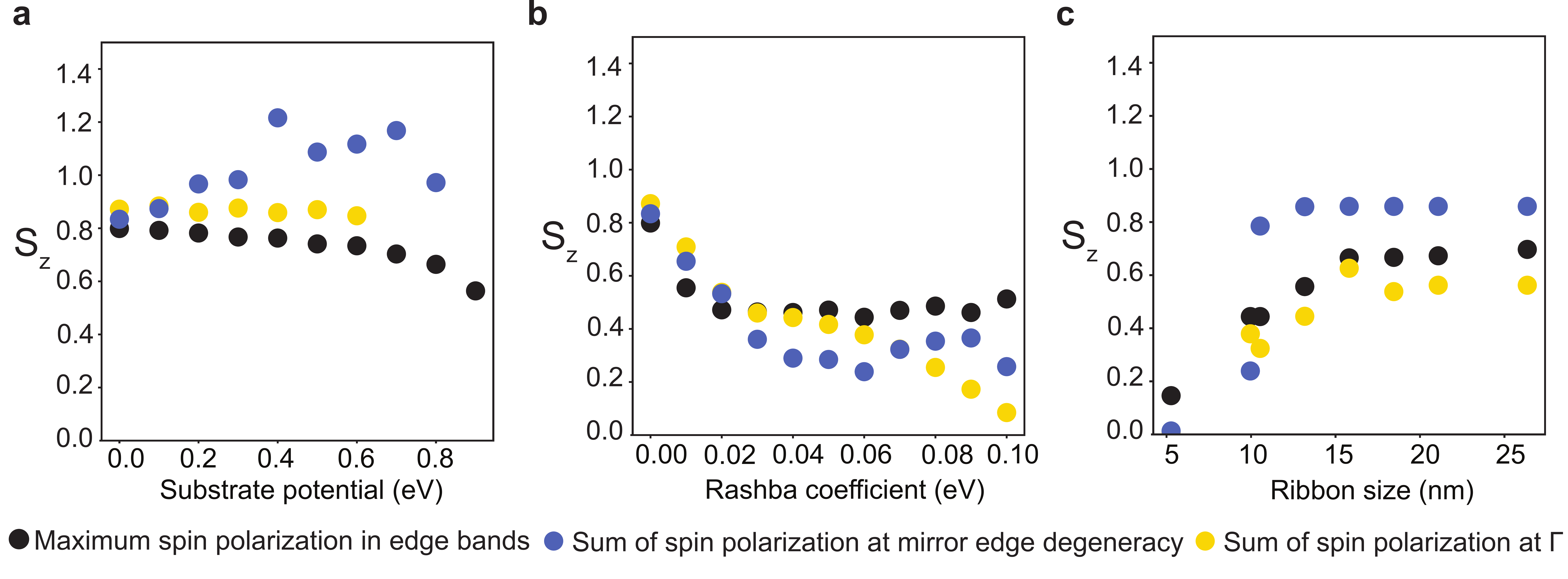}
  \end{center}
  \caption {\small Evolution of the $S_z$ spin polarization of the edge bands in armchair bismuthene nanoribbon as a function of (a) substrate potential, (b) external electric field and (c) the ribbon width. In (a) and (b), the ribbon width is kept at 9 nm. 
  In (c), a substrate potential of 0 eV and a field strength of $\lambda_R=0.06$ eV is used.
 }\label{fig8}
  \vspace{\baselineskip}
  \end{figure}
We have also investigated the effect of the substrate potential and the electric field on the total edge band $S_z$ polarization in armchair nanoribbons as shown in Figure \ref{fig8}.
It can be observed that, in general, the spin current becomes weaker as the perturbation due to substrate or external electric field becomes stronger (black circles in Figure \ref{fig8} (a) and (b)). 
The substrate potential enhances the spin current at the mirror edge degeneracy point DC5 and maintains the spin current at $\Gamma$. 
Compared to the substrate potential, the electric field causes a stronger loss in the spin current. 
In addition to the substrate potential and the external electric field, the width of the nanoribbon can also alter the spin accumulation at the edges. 
In Figure \ref{fig8} (c), we illustrate the evolution of $S_z$ spin polarization as a function of the width of the bismuthene armchair nanoribbon. Edge spin polarization is low in narrow nanoribbons, while it stabilizes for ribbons wider than 15 nm. The width of the nanoribbon has a contrasting influence compared to the substrate potential and external electric field---wide nanoribbons can reduce the loss of the spin polarization induced due to the substrate potential and external electric field.

 \begin{figure}[htbp]
  \begin{center}

  \includegraphics[scale=0.18]{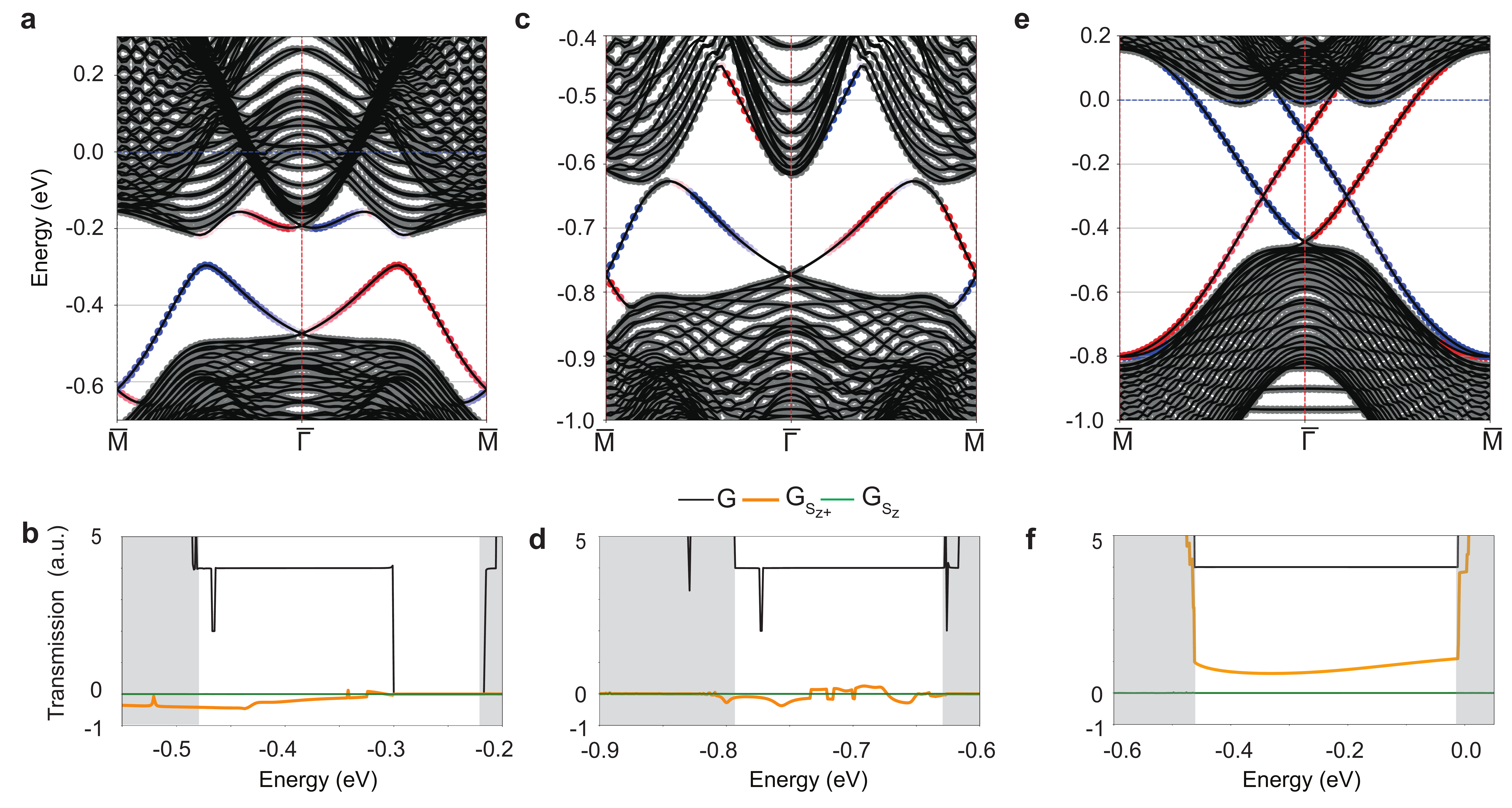}
  \end{center}
  \caption {\small Extracting unconventional spin segments and achieving spin transport transition in bismuthene armchair nanoribbons.  (a) and (c) show two distinct edge spin configurations achieved by using combination of the substrate effect, electric field and finite size of the nanoribbon. In (a), a substrate potential of 0.4 eV and a field strength of $\lambda_R=0.09$ eV is used, while in (c), a substrate potential of 0.9 eV and a field strength of $\lambda_R=0.08$ eV is used. The width of the nanoribbon is 15.8 nm in both cases. (b,d) Calculated transmission coefficients for the corresponding edge dispersions shown in (a,c). Change of the sign of transmission of spin-up electrons means the spin transport direction is reversed.   Grey shaded area represents the bulk region. (e) shows the edge band structure of armchair nanoribbon if a substrate potential of 0.8 eV is applied to $p_x$/$p_y$ states. The width of the nanoribbon is 15.8 nm. (f) shows the transmission corresponding to (e). The red (blue) circles in (a),(c) and (e) indicating spin-up (spin-down) projections in the bands with significant edge contributions, while the grey circles on bands represent contributions from bulk.
  }\label{fig9}
  \vspace{\baselineskip}
  \end{figure}

Figure \ref{fig9} demonstrates how combinations of control knobs, namely the substrate effect, electric field, and nanoribbon width, can lead to different modes of edge spin transport, as we designed in Figure \ref{fig6} (f). The first configuration in Figure \ref{fig9} (a) showcases the potential of bulk-edge interactions in isolating certain spin segments of interest. In this case, the original top edge branch is entirely buried in the bulk conduction bands, leaving only the original bottom half of the edge spin texture (as shown in Figure \ref{fig6} (c)). This approach could be useful for isolating spin channels from a highly entangled band dispersion. The transmission measurement shows that the transport direction of spin-up electrons is reversed, indicated by the negative transmission probability in Figure \ref{fig9} (b). This can be attributed to the larger spin polarization in the spin-flipped bands near $\overline{M}$ compared to the helical bands degenerate at $\overline{G}$.
The second example, shown in Figure \ref{fig9} (c), has electronic states at all energy levels, resulting in a fully conducting nanoribbon. In this case, the spin currents are much weaker compared to the gapped case in Figure \ref{fig9} (a). Nevertheless, we have successfully achieved a full unconventional spin texture in the conducting channel, as suggested in Figure \ref{fig6} (c) and (d), resembling the ideal configurations proposed in Figure \ref{fig6} (f). Figure \ref{fig9} (d) shows that the edge channels are still conducting and display a quantized conductance. The spin conductance exhibits a small and fluctuating pattern, suggesting that the spin transport is both energy and momentum dependent.
Regarding the selection of an appropriate substrate for experimental realization of this effect, we propose considering h-BN as a viable option due to its minimal misfit, comparable hexagonal structure, and weak interaction with bismuthene \cite{MAHESHWARAN2023108060}. Furthermore, the interaction between the substrate and bismuthene can be fine-tuned by applying external pressure \cite{10.1063/5.0123283}.

We can further demonstrate the impact of tuning bulk-edge interactions by downshifting the $p_x$/$p_y$ states instead of the $p_z$ states. By significantly changing the $p_x$/$p_y$ shift, we can eliminate spin flips in the pristine band structure and transform the edge band structure into a conventional helical state. Figure \ref{fig9} (e) shows the realization of this spin transport transition with a $p_x$/$p_y$ shift of 0.8 eV. The edge states exhibit a conventional helical spin texture, as seen by the uniform spin-up transmission probability in Figure \ref{fig9} (f).

Our investigation has uncovered novel spin transport patterns in topological materials, which challenges the conventional understanding of helical spin transport. We have shown that the spin-flipped channel near the mirror symmetry-protected edge degeneracy can significantly affect the spin transport direction and strength. This spin flip is momentum-dependent, which means it does not easily scatter with the other helical state. Although the net spin transport is zero in our simple two-terminal setting, we expect that this spin current can be revealed with proper multi-terminal setups, as demonstrated in previous investigations on helical systems \cite{PhysRevLett.95.226801}.
Finally, Our findings suggest that substrate potential and external fields can be delicately controlled to restore the spin flip to conventional helical transport. This suggests that we can use bulk-edge interactions to achieve spin transport transition without disrupting the bulk topology of the material. Overall, our investigation opens up new avenues for exploring and manipulating spin transport in topological materials.

\section{Conclusions}

In this study, we have demonstrated a comprehensive approach to manipulate the electronic properties of two-dimensional planar topological crystalline insulators (TCIs). By utilizing first principles calculations and symmetrized Wannier tight-binding models, we have explored the impact of various factors, including edge configurations, finite size effects, substrate effects, and external electric field, on the bulk-edge interactions. Our findings provide a complete toolkit to tailor the edge band dispersions and spin polarizations of TCIs by controlling the band alignments between the bulk and edge bands.
Furthermore, our results showcase the resilience of the pristine edge band dispersions and spin polarization characters to symmetry-breaking perturbations such as external electric fields. This resilience enables us to extract specific edge branches or spin channels with unconventional helical spin texture by using different combinations of tuning factors, which can form either gapped or conducting states along the edges of the armchair nanoribbon.
Our research opens up new opportunities for future experimental studies to manipulate the edge transport behavior of topological materials by controlling the bulk-edge interactions. Moreover, our results can be generalized to other two-dimensional topological thin films, providing a framework for achieving flexible and reversible control of the edge currents in topological transistors. 

\section{Acknowledgements}

Authors acknowledge the support of the Australian Research Council’s Centre of Excellence in Future Low-Energy Electronics Technologies (CE170100039). The authors also acknowledge the computational support from
Australian National Computing Infrastructure (NCI), the NCI Adapter Scheme and Pawsey Supercomputing Centre. 



\bibliographystyle{model1a-num-names}
\bibliography{ref}

\end{document}